\documentclass[prl, twocolumn, nobalancelastpage, showpacs]{revtex4}

\usepackage[]{graphicx}
\usepackage{amsfonts}

\newcommand \ket[1] {| #1 \rangle}

\begin{document}

\title{Description of a quantum convolutional code}

\author{Harold Ollivier} 
\affiliation{INRIA, Projet CODES, BP 105, F-78153 Le Chesnay, France} 
\author{Jean-Pierre Tillich}
\affiliation{INRIA, Projet CODES, BP 105, F-78153 Le Chesnay, France}



\begin{abstract}
We describe a quantum error correction scheme aimed at protecting a
flow of quantum information over long distance communication. It is
largely inspired by the theory of classical convolutional codes which
are used in similar circumstances in classical communication. The
particular example shown here uses the stabilizer formalism, which
provides an explicit encoding circuit. An associated error estimation
algorithm is given explicitly and shown to provide the most likely
error over {\em any} memoryless quantum channel, while its complexity
grows only linearly with the number of encoded qubits.
\end{abstract}

\maketitle

In recent years, the discovery and development of quantum computation
and communication has shed new light on quantum physics. The potential
applications of these new fields encompass a wide variety of subjects,
ranging from unconditionally secure secret key generation
protocols~\cite{BB84a} to efficient integer factoring
algorithms~\cite{Sho97a} or enhancement of communication
complexity~\cite{BCW98a}. However, the practical realization of such
protocols and algorithms remains a very involved task mainly because
of the inherent instability of quantum superpositions~\cite{Zur91a} as
well as intrinsic imprecisions of the physical devices that process
quantum information. These errors wipe out the quantum superpositions
together with entanglement, which are usually seen as key resources of
the power of quantum algorithms and protocols~\cite{JL02a}. Hence,
protecting the quantum nature of information became one of the most
important challenges to prove the feasibility of quantum
computers. The discovery of quantum error correction
schemes~\cite{Sho95a, LMPZ96a} notably opened the future of large
scale quantum information processing: a certain, but unfortunately
very small, degree of imprecision can be tolerated at each step of a
quantum transformation and still allow a speed-up over classical
information processing~\cite{AB96a, Zal96a}. However, building a
fault-tolerant quantum computer remains largely out of reach of the
present day practical realizations, principally because of the large
number of physical qubits required to account for the error
correction.

On the other hand, quantum cryptography and more generally the field
of quantum communication seems more promising in a near future. Some
quantum key distribution protocols have been implemented and the
associated devices seem to be close to
commercialization~\cite{SGGR02a}. Within this context, we construct a
new family of codes ---~quantum convolutional codes~--- aimed at
protecting a stream of quantum information in a long distance
communication. They are the correct generalization to the quantum
domain of their classical analogues, and hence inherit their most
important properties. First, they have a {\em maximum likelihood}
error estimation algorithm for {\em all} memoryless channels with a
complexity growing {\em linearly} with the number of encoded
qubits. This is an important issue since finding the most likely error
---~a strategy which allows to determine the most likely sent
codeword~--- is in general a hard task: for a generic family of block
codes with constant rate, the maximum likelihood error estimation
algorithm has a complexity growing {\em exponentially} with the number
of encoded qubits. Hence, generic block codes rapidly require to
employ suboptimal error estimation procedures which, as a consequence,
do not exploit the whole error correcting capabilities of the
code. Moreover, our algorithm can easily handle variations in the
properties of the communication channel (i.e.~a change in the single
qubit error probabilities). The second advantage of quantum
convolutional codes is their ability to perform the encoding of the
qubits {\em online} (i.e.\@~as they arrive in the encoder). Thus, it
is not necessary to wait for all the qubits to be ready to start
sending the encoded state through the communication channel: it
reduces the overall processing time of the qubits which is an
additional source of decoherence. Note that an attempt at defining
quantum convolutional codes has been made some time ago~\cite{Cha98a,
Cha98b}, but missed some crucial points concerning the error
estimation algorithm as well as error propagation properties.

In this letter, we deal with a specific example drawn from our general
theory. We construct a quantum convolutional code achieving a rate
equal to $1/5$: we explain how to encode and decode a stream of qubits
efficiently, and we expose the maximum likelihood error estimation
algorithm. This will give all the necessary intuition to understand
how to generalize the present results to a wider
framework~\cite{OT03b}.

\paragraph{Description of the code~---}
The particular code we wish to present is best described by using the
stabilizer formalism~\cite{Got97a}. This provides a simple way to
understand the encoding and decoding operations. Moreover, the error
syndromes can be easily identified, which considerably simplifies the
description of the error estimation algorithm. We use the following
standard notations for the Pauli operators acting on a single qubit:
\begin{equation} 
X = \left (\begin{array}{cc} 0 & 1 \\ 1 & 0 \end{array}\right),\; Y =
\left (\begin{array}{cc} 0 & i \\ -i & 0 \end{array}\right),\; Z =
\left (\begin{array}{cc} 1 & 0 \\ 0 & -1
\end{array}\right), \end{equation} 
so that $ZX=Y$. The identity matrix will be denoted by $I$. Since
convolutional codes are designed to deal with a stream of information
qubits, the number of generators of the stabilizer group will possibly
be infinite. However in practice, transmission starts and ends at a
given time, which means that we only consider generators made of a
finite number of Pauli operators.

The code subspace is described by the generators of its stabilizer
group, $S$. These generators are given by:
\begin{equation} 
\begin{array}{lcp{1em}@{}p{1em}@{}p{1em}@{}p{1em}@{}p{1em}@{}p{1em}@{}p{1em}@{}p{1em}@{}l} 
M_0 & = & $X$&$Z$&$I$&$I$&$I$&$I$&$I$&$I$&$\ldots$, \\
M_1 & = & $Z$&$X$&$X$&$Z$&$I$&$I$&$I$&$I$&$\ldots$, \\ 
M_2 & = & $I$&$Z$&$X$&$X$&$Z$&$I$&$I$&$I$&$\ldots$, \\ 
M_3 & = & $I$&$I$&$Z$&$X$&$X$&$Z$&$I$&$I$&$\ldots$, \\
M_4 & = & $I$&$I$&$I$&$Z$&$X$&$X$&$Z$&$I$&$\ldots$, \\
M_{4i+j} & = & \multicolumn{9}{l}{I^{\otimes 5i}\otimes M_j, \ 0 < i, \ 1 \leq j \leq 4 },\\ 
M_{\infty} & = & \multicolumn{2}{r}{\ldots}&$I$&$I$&$I$&$I$&$Z$&$X$&.
\end{array}
\end{equation}
It is easy to check that all the generators commute and are
independent. Thus, the code subspace (i.e.\@~the largest common
eigenspace of the generators with eigenvalue $+1$) is non-trivial.

An important point to address when considering stabilizer codes is the
ability to manipulate encoded information. Namely, we want to find the
encoded Pauli operators $\overline X_i$, $\overline Z_i$ corresponding
to logical qubit $i$. These operators must satisfy the following
relations:
\begin{eqnarray} 
&& \overline X_i, \, \overline Z_i \in  N(S) - S, \label{eq:condition1}\\
&&{[\overline X_i, \overline X_j]}  =  0, \label{eq:condition2}\\
&&{[\overline Z_i, \overline Z_j]}  =  0, \label{eq:condition3}\\
&&{[\overline X_i, \overline Z_j]}  =  0, \ i \neq j , \label{eq:condition4}
\end{eqnarray}
where $N(S)$ denotes the normalizer of
$S$. Equation~(\ref{eq:condition1}) states that the encoded Pauli
operators leave the code subspace globally invariant, but have a
non-trivial action on its elements, while the
Equations~(\ref{eq:condition2}-\ref{eq:condition4}) ensure that
manipulating qubit $i$ does not affect other qubits. There exists a
great choice of different sets of such operators, however they are not
equivalent in the perspective of effectively manipulating the encoded
quantum information in an easy way: in practice only those with a
small number of terms different from the identity are useful. For our
particular example, such set exists and has a structure invariant by a
shift of five qubits:
\begin{equation}
\begin{array}{lccccccccc}
\overline X_1 &=& I&Z&I&X&I&Z&I&I\ldots,\\
\overline Z_1 &=& I&Z&Z&Z&Z&Z&I&I\ldots,\\
\overline X_n &=& \multicolumn{8}{l}{I^{\otimes 5n}\otimes \overline X_1,\ n > 1,} \\ 
\overline Z_n &=& \multicolumn{8}{l}{I^{\otimes 5n}\otimes \overline Z_1,\ n > 1.} \end{array}
\end{equation}
Hence, a unitary transformation on a single encoded qubit will in
general be implemented by a unitary transformation on five physical
qubits. 

At this point, one can wonder what in this code differs from a generic
block code. The answer to this question comes from the particular
structure of the stabilizer generators: beside $M_0$ and $M_\infty$,
the generators of the stabilizer group can be casted into sets of
constant size (e.g.~four), each set acting on a fixed number
(e.g.~seven) of consecutive qubits. In addition, each set has a fixed
overlap (e.g.~of two qubits) with the set immediately before and
immediately after. This very peculiar structure defines quantum
convolutional codes and we can prove~\cite{OT03b} that this implies
the possibility of online encoding and the existence of an efficient
error estimation algorithm.

\paragraph{Encoding circuit~---}
As explained in D.\@~Gottesman's Ph.D.~thesis~\cite{Got97a}, there are
various ways to realize the encoding into the code subspace. However,
for convolutional codes, they are not equivalent: standard encoding
circuits usually require to wait until the last `to-be-protected'
qubit has been obtained before sending the encoded state. In this
section, we explain how to take advantage of the structure of the
stabilizer generators to overcome this limitation and encode the
qubits {\em online}. We first exhibit a map from the computational
basis of the `to-be-protected' qubits to a basis of the code
subspace. As a second step, we derive the quantum circuit implementing
this map in a unitary way.

More precisely, consider the following set of states:
\begin{eqnarray}
&& \left \{ \ket{\psi(c_1,c_2,c_3,\ldots)}\right \}_{c_i \in \{0,1\}} = \\ 
&&  \left \{ P  \ket{0,0,0,0,0,c_1,0,0,0,0,c_2,0,0,0,0,c_3, \ldots} \right\}_{c_i \in \{0,1\}}, \nonumber 
\end{eqnarray}
where $P = \prod_i (I + M_i)/\sqrt 2$ is the projection operator onto
the code subspace. Since $\overline Z_i$ commutes with all the
generators of the stabilizer group, the following equation holds for
any element of the set:
\begin{eqnarray}
&&\overline Z_i P \ket{0,0,0,0,0,c_1,0,0,0,0,c_2,0,0,0,0,c_3, \ldots} = \\
&&\ (-1)^{c_i} P \ket{0,0,0,0,0,c_1,0,0,0,0,c_2,0,0,0,0,c_3, \ldots}. \nonumber
\end{eqnarray}
This implies that $\{ \ket{\psi(c_1,c_2,c_3,\ldots)}\}_{c_i \in
\{0,1\}}$ is an orthonormal basis of the code subspace. Hence, the
natural encoding consists in mapping the computational basis of the
`to-be-protected' qubits, $\{\ket{c_1,c_2,c_3,\ldots}\}_{c_i \in
\{0,1\}}$ into the basis $\{ \ket{\psi(c_1,c_2,c_3,\ldots)}\}_{c_i \in
\{0,1\}}$.

In practice, to encode a stream of qubits $q_i$, we first add to it
ancillary qubits in the $\ket 0$ state such that the `to-be-protected'
qubit $i$ is now at the position $5i+1$. Then, we need to implement
$P$ for these specific input states as a unitary transformation onto
the whole Hilbert space. This can be done in full generality as
explained in~\cite{Got97a}, and gives the encoding circuit of
\textsc{Fig}.~\ref{fig:encoding1}. From this simple example, it is
easy to understand that the possibility of online encoding for quantum
convolutional codes is a consequence of the finite extension of the
support of the generators of the stabilizer group and of the encoded
Pauli operators. Also note that alternative encoding methods can be
found and can be relevant when considering some specific applications,
but these issues are beyond the scope of this letter.

\paragraph{Error propagation and online decoding~---} 
Due to their very specific nature, convolutional codes propagate
information contained in a given qubit to its successors (see again
\textsc{Fig}.~\ref{fig:encoding1}). During the decoding process
(i.e.~the inverse of encoding) this can actually become a problem: an
error affecting a {\em finite} number of qubits before decoding can
propagate through the decoding circuit and finally affect an {\em
infinite} number of qubits.  Such error is called {\em catastrophic}.
It is worth mentioning that this issue is to the quantum domain:
classical convolutional encoders might have catastrophic
errors~\cite{JZ99a, Lee97a}. Fortunately, in both cases,
non-catastrophic encoders exist. More precisely, given a specific
encoder one can find a procedure to determine whether it has
catastrophic errors or not. For classical codes this is a well known
result established by Massey and Sain~\cite{SM68a}. For the quantum
domain, we can show~\cite{OT03b} that the circuit of non-catastrophic
encoders fulfills the following requirement: its gates form a finite
number of layers and commute with each other inside a layer. The idea
behind this theorem is simple. In general, an error affecting some
qubits will propagate to all the other qubits involved in a gate with
the erroneous ones. When those qubits are further used in other gates
the error continues to propagate until no more gates are applied. The
commutation relation together with the finiteness of the number of
layers ensures that, for any finite size error, only a finite number
of gates will enter in the propagation process. Thus all errors are
non-catastrophic. \textsc{Fig}.~\ref{fig:decoding1} illustrates this
`pearl-necklace' structure for our example, and thus proves that our
rate $1/5$ quantum convolutional code is non-catastrophic.

Moreover, it can be shown that this condition implies the existence of
a {\em forward} decoding scheme: there is no need to wait for the last
qubit to start decoding (see \textsc{Fig}.~\ref{fig:decoding1}). For
non-catastrophic codes, both encoding and decoding can be done
online~\cite{OT03b}.

\paragraph{Maximum likelihood error estimation~---}
An error correcting code aims at protecting information sent over a
noisy communication channel by letting the receiver infer which error
possibly affected the information. This is the role of the error
estimation algorithm. On average, the correct information is most
often retrieved when the estimated error coincides with the most
likely error. Thus, it is both of theoretical and practical relevance
to have an efficient maximum likelihood error estimation algorithm for
our quantum convolutional codes. In this section, we exhibit such
algorithm. It is indeed the quantum analogue of the well-known Viterbi
algorithm for classical convolutional codes. The Viterbi algorithm
realizes a maximum likelihood error estimation on all memoryless
channels with a complexity linear in the number of encoded bits. This
explains why classical convolutional codes are so widely used for
reducing the noise on communication channels.

Our algorithm for quantum convolutional codes processes the
information obtained through the syndrome in order to infer the most
likely error. The circuit for obtaining the syndromes follows the
usual phase estimation scheme: an ancillary qubit is prepared in the
$\ket 0$ state; undergoes a Hadamard transform; controls the
application of one of the generator $M_i$ of the stabilizer group;
again undergoes a Hadamard transform; and is finally measured in the
$\{\ket 0, \ket 1\}$ basis. Then, the algorithm updates a list of
maximum likelihood error candidates by looking at a small number of
syndromes at a time, and by taking local decisions. It is preceded and
followed by appropriate initialization and termination steps.

The initialization step lists all error candidates, $\{E_j^0\}_j$, for
the first two qubits which are compatible with the syndrome
$M_0$. There are exactly $8 = 4^2/2$ of them (there are $4^2$
different operators with support on the first two qubits, but the
constraint associated with $M_0$ divides this set into two equal
parts). This list constitutes the input of the main loop of the
algorithm. At step $i$, the algorithm constructs a list of some most
likely error candidates, $\{E_j^i\}_j$ compatible with the syndromes
$M_0$ to $M_{4i}$. Each candidate $E_j^i$ is thus specified only on
qubits $1$ to $5i+2$. The crucial point of the algorithm is to
maintain a fixed size of this list, and hence to avoid the exponential
blow up that would arise when listing all error candidates compatible
with these syndromes. More precisely, $E_j^i$ is a most likely
candidate whose restriction on qubit $5i+1$ and $5i+2$ is prescribed
by the index $j$ running over the set of $16$ possible errors
affecting those two qubits. The computation of any error candidate
$E_k^{i+1}$ is easily achieved provided $\{E_j^{i}\}_j$: consider the
set of all possible extensions of the error candidates $E_j^i$ to
qubit $5i+3$ to $5i+7$ with the prescribed error $k$ at position
$5i+6$ and $5i+7$. It is easy to check that any such element is now
compatible with syndromes $M_0$ to $M_{4(i+1)}$. The specific
candidate $E_k^{i+1}$ is chosen to be the most likely operator among
the elements of the latter set (in case of tie, one is chosen at
random). This procedure is continued until reaching $M_\infty$, which
again selects half of the candidates. The termination of the algorithm
outputs the most likely candidate among the remaining ones. This
constitutes the most likely error given the value of all the syndromes
for the received stream of qubits~\cite{OT03b}.

The main property used to prove this fact is related to the structure
of the generators of the stabilizer group: the value of the syndromes
associated to $M_{4i+1}$ to $M_{4i+4}$ depend on the syndromes $M_0$
to $M_{4i}$ only through the error operators at position $5i+1$ and
$5i+2$. Thus, taking a sequence of local decisions allows to construct
a list of error candidates among which one will coincide with the most
likely error until qubit $5i+2$ while maintaining a linear complexity
of the algorithm as the number of encoded qubits increases. Note that,
the error maximizing the likelihood is known when the last syndrome is
measured. Hence, it is in principle necessary to wait till the end of
the transmission to actually correct the estimated error. However, as
for the classical Viterbi algorithm, numerical simulations show that
the different candidates at a given step coincide with the most likely
error except on their last few positions. Thus, in practice it is
possible to estimate the error online. In addition, we want to stress,
that without increasing its complexity, this algorithm can take into
account all memoryless quantum channels even if the single qubit error
probabilities are not constant in time. For example, one could imagine
that the qubits are photons sent through an optical fiber, and that
the probabilities are evaluated by sending probe photons containing no
useful information. Finally, as the codes described here are the exact
translation to the quantum setting of the classical convolutional
codes, one can also derive suboptimal error estimation algorithms (for
their classical analogues see~\cite{JZ99a, Lee97a}). Most importantly,
quantum convolutional codes can be decoded iteratively and allow
quantum turbo decoding~\cite{OT03b}.

\paragraph{Conclusion~---}
In this letter, we presented the theory of quantum convolutional codes
by an example. We gave explicitly the associated encoding and decoding
circuits, as well as a low complexity maximum likelihood error
estimation algorithm. We believe that such codes could be used to
reduce errors for long distance quantum communications provided that
we are able to perform a small and fixed number quantum gates with
good fidelity. Moreover, the tools developed for quantum convolutional
codes can be used to translate other families of classical codes to
the quantum domain, like for instance low density parity check codes.

\begin{acknowledgments}
Part of this work was done when H.O.\@ was visiting the Perimeter
Institute and the Institute for Quantum Computing in Waterloo,
Canada. Useful discussions with J.~Kempe, R.~Laflamme and D.~Poulin
are gratefully acknowledged.
\end{acknowledgments}


\begin{figure}[htbp]

\begin{picture}(195,135)

\multiput(0,130)(000,-10){13}{\line(1,0){10}} 

\multiput(18,130)(000,-10){13}{\line(1,0){7}} 

\multiput(10,106)(0,-10){4}{\framebox(8,8){H}}
\multiput(10,56)(0,-10){4}{\framebox(8,8){H}}
\multiput(10,126)(0,-10){1}{\framebox(8,8){H}}
\multiput(10,6)(0,-10){1}{\framebox(8,8){H}}
\multiput(10,120)(000,-50){3}{\line(1,0){8}} 

\multiput(-12,126)(0,-10){6}{\makebox(8,8)[l]{$\ket 0$}}
\multiput(-12,56)(0,-10){4}{\makebox(8,8)[l]{$\ket 0$}}
\put(-12,66){\makebox(8,8)[l]{$q_1$}}
\put(-12,16){\makebox(8,8)[l]{$q_2$}}
\put(-12,6){\makebox(8,8)[l]{$\ket 0$}}

\put(25,130){\line(1,0){8}}
\multiput(25,110)(000,-10){11}{\line(1,0){8}} 
\put(25,116){\framebox(8,8){Z}}
\put(29,130){\line(0,-1){6}}
\put(29,130){\circle*{3}}

\multiput(33,130)(000,-10){13}{\line(1,0){7}} 

\multiput(40,90)(000,-10){9}{\line(1,0){8}} 
\put(40,126){\framebox(8,8){Z}}
\put(40,116){\framebox(8,8){X}}
\put(40,110){\line(1,0){8}}
\put(44,110){\circle*{3}}
\put(40,100){\line(1,0){8}}
\put(44,126){\line(0,-1){2}}
\put(44,116){\line(0,-1){6}}

\multiput(48,130)(000,-10){13}{\line(1,0){7}} 

\multiput(55,80)(000,-10){8}{\line(1,0){8}} 
\put(55,130){\line(1,0){8}}
\put(55,116){\framebox(8,8){Z}}
\put(55,106){\framebox(8,8){X}}
\put(55,100){\line(1,0){8}}
\put(59,100){\circle*{3}}
\put(55,90){\line(1,0){8}}
\put(59,116){\line(0,-1){2}}
\put(59,106){\line(0,-1){6}}

\multiput(63,130)(000,-10){13}{\line(1,0){7}} 

\multiput(70, 70)(000,-10){7}{\line(1,0){8}} 
\multiput(70,130)(0,-10){2}{\line(1,0){8}}
\put(70,106){\framebox(8,8){Z}}
\put(70,96){\framebox(8,8){X}}
\put(70,90){\line(1,0){8}}
\put(74,90){\circle*{3}}
\put(70,80){\line(1,0){8}}
\put(74,106){\line(0,-1){2}}
\put(74,96){\line(0,-1){6}}

\multiput(78,130)(000,-10){13}{\line(1,0){7}} 

\multiput(85, 60)(000,-10){6}{\line(1,0){8}} 
\multiput(85,130)(0,-10){3}{\line(1,0){8}}
\put(85,96){\framebox(8,8){Z}}
\put(85,86){\framebox(8,8){X}}
\put(85,80){\line(1,0){8}}
\put(89,80){\circle*{3}}
\put(85,66){\framebox(8,8){Z}}
\put(89,96){\line(0,-1){2}}
\put(89,86){\line(0,-1){12}}

\multiput(93,130)(000,-10){13}{\line(1,0){7}} 

\multiput(100, 40)(000,-10){4}{\line(1,0){8}} 
\multiput(100,130)(0,-10){5}{\line(1,0){8}}
\put(100,76){\framebox(8,8){Z}}
\put(100,66){\framebox(8,8){X}}
\put(100,60){\line(1,0){8}}
\put(104,60){\circle*{3}}
\put(100,50){\line(1,0){8}}
\put(104,76){\line(0,-1){2}}
\put(104,66){\line(0,-1){6}}

\multiput(108,130)(000,-10){13}{\line(1,0){7}} 

\multiput(115, 30)(000,-10){3}{\line(1,0){8}} 
\multiput(115,130)(0,-10){6}{\line(1,0){8}}
\put(115,66){\framebox(8,8){Z}}
\put(115,56){\framebox(8,8){X}}
\put(115,50){\line(1,0){8}}
\put(119,50){\circle*{3}}
\put(115,40){\line(1,0){8}}
\put(119,66){\line(0,-1){2}}
\put(119,56){\line(0,-1){6}}

\multiput(123,130)(000,-10){13}{\line(1,0){7}} 

\multiput(130, 20)(000,-10){2}{\line(1,0){8}} 
\multiput(130,130)(0,-10){7}{\line(1,0){8}}
\put(130,56){\framebox(8,8){Z}}
\put(130,46){\framebox(8,8){X}}
\put(130,40){\line(1,0){8}}
\put(134,40){\circle*{3}}
\put(130,30){\line(1,0){8}}
\put(134,56){\line(0,-1){2}}
\put(134,46){\line(0,-1){6}}

\multiput(138,130)(000,-10){13}{\line(1,0){7}} 

\multiput(145, 10)(000,-10){1}{\line(1,0){8}} 
\multiput(145,130)(0,-10){8}{\line(1,0){8}}
\put(145,46){\framebox(8,8){Z}}
\put(145,36){\framebox(8,8){X}}
\put(145,30){\line(1,0){8}}
\put(149,30){\circle*{3}}
\put(145,16){\framebox(8,8){Z}}
\put(149,46){\line(0,-1){2}}
\put(149,36){\line(0,-1){12}}

\multiput(153,130)(000,-10){13}{\line(1,0){7}} 

\multiput(160,130)(0,-10){10}{\line(1,0){8}}
\put(160,26){\framebox(8,8){Z}}
\put(160,16){\framebox(8,8){X}}
\put(160,10){\line(1,0){8}}
\put(164,10){\circle*{3}}
\put(164,26){\line(0,-1){2}}
\put(164,16){\line(0,-1){6}}

\multiput(168,130)(000,-10){13}{\line(1,0){7}} 

\multiput(175,130)(0,-10){11}{\line(1,0){8}}
\put(175,16){\framebox(8,8){Z}}
\put(175,6){\framebox(8,8){X}}
\put(179,16){\line(0,-1){2}}
\put(179,6){\line(0,-1){2}}

\multiput(183,130)(000,-10){13}{\line(1,0){7}} 

\multiput(190,130)(0,-10){12}{\line(1,0){8}}
\put(190,6){\framebox(8,8){Z}}
\put(194,6){\line(0,-1){2}}

\multiput(198,130)(0,-10){13}{\line(1,0){10}}

\end{picture}

\caption{Beginning of the circuit realizing the encoding once the
ancillary qubits have been added to the stream containing the initial
quantum information (qubits $q_1, q_2, \ldots$). $H$ is the Hadamard
transform, and the dot represents the control qubit for a given
gate. The circuit is run from left to right. When all the
transformations have been performed for a given qubit, it can be sent
through the communication channel.}\label{fig:encoding1}
\end{figure}
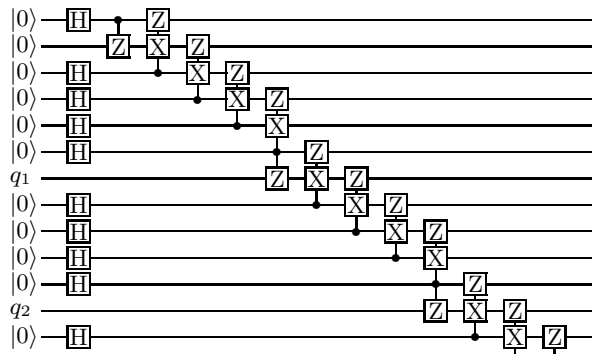

\begin{figure}[htbp]
\begin{picture}(210,135)

\multiput(0,130)(000,-10){13}{\line(1,0){10}} 

\multiput(18,130)(000,-10){13}{\line(1,0){7}} 

\multiput(10,106)(0,-10){4}{\framebox(8,8){H}}
\multiput(10,56)(0,-10){4}{\framebox(8,8){H}}
\multiput(10,126)(0,-10){1}{\framebox(8,8){H}}
\multiput(10,6)(0,-10){1}{\framebox(8,8){H}}
\multiput(10,120)(000,-50){3}{\line(1,0){8}} 

\multiput(-12,126)(0,-10){6}{\makebox(8,8)[l]{$\ket 0$}}
\multiput(-12,56)(0,-10){4}{\makebox(8,8)[l]{$\ket 0$}}
\put(-12,66){\makebox(8,8)[l]{$q_1$}}
\put(-12,16){\makebox(8,8)[l]{$q_2$}}
\put(-12,6){\makebox(8,8)[l]{$\ket 0$}}

\put(25,130){\line(1,0){8}}
\put(25,116){\framebox(8,8){Z}}
\put(29,130){\line(0,-1){6}}
\put(29,130){\circle*{3}}

\put(29,80){\circle*{3}}
\put(29,80){\line(0,-1){16}}
\put(25,56){\framebox(8,8){Z}}

\put(29,30){\circle*{3}}
\put(29,30){\line(0,-1){16}}
\put(25,6){\framebox(8,8){Z}}

\multiput(25,20)(0,10){4}{\line(1,0){8}}
\multiput(25,70)(0,10){5}{\line(1,0){8}}

\multiput(33,130)(000,-10){13}{\line(1,0){7}} 

\put(40,126){\framebox(8,8){Z}}
\put(40,116){\framebox(8,8){X}}
\put(40,110){\line(1,0){8}}
\put(44,110){\circle*{3}}
\put(40,100){\line(1,0){8}}
\put(44,126){\line(0,-1){2}}
\put(44,116){\line(0,-1){6}}

\put(40,90){\line(1,0){8}}
\put(40,40){\line(1,0){8}} 

\put(40,76){\framebox(8,8){Z}}
\put(40,66){\framebox(8,8){X}}
\put(40,60){\line(1,0){8}}
\put(44,60){\circle*{3}}
\put(40,50){\line(1,0){8}}
\put(44,76){\line(0,-1){2}}
\put(44,66){\line(0,-1){6}}

\put(40,26){\framebox(8,8){Z}}
\put(40,16){\framebox(8,8){X}}
\put(40,10){\line(1,0){8}}
\put(44,10){\circle*{3}}
\put(44,26){\line(0,-1){2}}
\put(44,16){\line(0,-1){6}}

\multiput(48,130)(000,-10){13}{\line(1,0){7}} 

\put(55,130){\line(1,0){8}}
\put(55,116){\framebox(8,8){Z}}
\put(55,106){\framebox(8,8){X}}
\put(55,100){\line(1,0){8}}
\put(59,100){\circle*{3}}
\put(55,90){\line(1,0){8}}
\put(59,116){\line(0,-1){2}}
\put(59,106){\line(0,-1){6}}

\put(55,80){\line(1,0){8}}
\put(55,30){\line(1,0){8}}

\put(55,66){\framebox(8,8){Z}}
\put(55,56){\framebox(8,8){X}}
\put(55,50){\line(1,0){8}}
\put(59,50){\circle*{3}}
\put(55,40){\line(1,0){8}}
\put(59,66){\line(0,-1){2}}
\put(59,56){\line(0,-1){6}}

\put(55,16){\framebox(8,8){Z}}
\put(55,6){\framebox(8,8){X}}
\put(59,16){\line(0,-1){2}}
\put(59,6){\line(0,-1){2}}

\multiput(63,130)(000,-10){13}{\line(1,0){7}} 

\multiput(70,130)(0,-10){2}{\line(1,0){8}}
\put(70,106){\framebox(8,8){Z}}
\put(70,96){\framebox(8,8){X}}
\put(70,90){\line(1,0){8}}
\put(74,90){\circle*{3}}
\put(70,80){\line(1,0){8}}
\put(74,106){\line(0,-1){2}}
\put(74,96){\line(0,-1){6}}

\put(70,70){\line(1,0){8}}
\put(70,20){\line(1,0){8}}

\put(70,56){\framebox(8,8){Z}}
\put(70,46){\framebox(8,8){X}}
\put(70,40){\line(1,0){8}}
\put(74,40){\circle*{3}}
\put(70,30){\line(1,0){8}}
\put(74,56){\line(0,-1){2}}
\put(74,46){\line(0,-1){6}}

\put(70,6){\framebox(8,8){Z}}
\put(74,6){\line(0,-1){2}}

\multiput(78,130)(000,-10){13}{\line(1,0){7}} 

\multiput(85,130)(0,-10){3}{\line(1,0){8}}
\put(85,96){\framebox(8,8){Z}}
\put(85,86){\framebox(8,8){X}}
\put(85,80){\line(1,0){8}}
\put(89,80){\circle*{3}}
\put(85,66){\framebox(8,8){Z}}
\put(89,96){\line(0,-1){2}}
\put(89,86){\line(0,-1){12}}

\put(85,60){\line(1,0){8}}
\put(85,10){\line(1,0){8}}

\put(85,46){\framebox(8,8){Z}}
\put(85,36){\framebox(8,8){X}}
\put(85,30){\line(1,0){8}}
\put(89,30){\circle*{3}}
\put(85,16){\framebox(8,8){Z}}
\put(89,46){\line(0,-1){2}}
\put(89,36){\line(0,-1){12}}

\multiput(93,130)(000,-10){13}{\line(1,0){10}} 


\multiput(225,130)(000,-10){13}{\line(-1,0){10}} 

\multiput(207,130)(000,-10){13}{\line(-1,0){7}} 

\multiput(207,106)(0,-10){4}{\framebox(8,8){H}}
\multiput(207,56)(0,-10){4}{\framebox(8,8){H}}
\multiput(207,126)(0,-10){1}{\framebox(8,8){H}}
\multiput(207,6)(0,-10){1}{\framebox(8,8){H}}
\multiput(207,120)(000,-50){3}{\line(1,0){8}} 

\put(200,130){\line(-1,0){8}}
\put(192,116){\framebox(8,8){Z}}
\put(196,130){\line(0,-1){6}}
\put(196,130){\circle*{3}}

\put(196,80){\circle*{3}}
\put(196,80){\line(0,-1){16}}
\put(192,56){\framebox(8,8){Z}}

\put(196,30){\circle*{3}}
\put(196,30){\line(0,-1){16}}
\put(192,6){\framebox(8,8){Z}}

\multiput(200,20)(0,10){4}{\line(-1,0){8}}
\multiput(200,70)(0,10){5}{\line(-1,0){8}}

\multiput(192,130)(000,-10){13}{\line(-1,0){7}} 

\put(177,126){\framebox(8,8){Z}}
\put(177,116){\framebox(8,8){X}}
\put(177,110){\line(1,0){8}}
\put(181,110){\circle*{3}}
\put(177,100){\line(1,0){8}}
\put(181,126){\line(0,-1){2}}
\put(181,116){\line(0,-1){6}}

\put(177,90){\line(1,0){8}}
\put(177,40){\line(1,0){8}} 

\put(177,76){\framebox(8,8){Z}}
\put(177,66){\framebox(8,8){X}}
\put(177,60){\line(1,0){8}}
\put(181,60){\circle*{3}}
\put(177,50){\line(1,0){8}}
\put(181,76){\line(0,-1){2}}
\put(181,66){\line(0,-1){6}}

\put(177,26){\framebox(8,8){Z}}
\put(177,16){\framebox(8,8){X}}
\put(177,10){\line(1,0){8}}
\put(181,10){\circle*{3}}
\put(181,26){\line(0,-1){2}}
\put(181,16){\line(0,-1){6}}

\multiput(177,130)(000,-10){13}{\line(-1,0){7}} 

\put(162,130){\line(1,0){8}}
\put(162,116){\framebox(8,8){Z}}
\put(162,106){\framebox(8,8){X}}
\put(162,100){\line(1,0){8}}
\put(166,100){\circle*{3}}
\put(162,90){\line(1,0){8}}
\put(166,116){\line(0,-1){2}}
\put(166,106){\line(0,-1){6}}

\put(162,80){\line(1,0){8}}
\put(162,30){\line(1,0){8}}

\put(162,66){\framebox(8,8){Z}}
\put(162,56){\framebox(8,8){X}}
\put(162,50){\line(1,0){8}}
\put(166,50){\circle*{3}}
\put(162,40){\line(1,0){8}}
\put(166,66){\line(0,-1){2}}
\put(166,56){\line(0,-1){6}}

\put(162,16){\framebox(8,8){Z}}
\put(162,6){\framebox(8,8){X}}
\put(166,16){\line(0,-1){2}}
\put(166,6){\line(0,-1){2}}

\multiput(162,130)(000,-10){13}{\line(-1,0){7}} 

\multiput(147,130)(0,-10){2}{\line(1,0){8}}
\put(147,106){\framebox(8,8){Z}}
\put(147,96){\framebox(8,8){X}}
\put(147,90){\line(1,0){8}}
\put(151,90){\circle*{3}}
\put(147,80){\line(1,0){8}}
\put(151,106){\line(0,-1){2}}
\put(151,96){\line(0,-1){6}}

\put(147,70){\line(1,0){8}}
\put(147,20){\line(1,0){8}}

\put(147,56){\framebox(8,8){Z}}
\put(147,46){\framebox(8,8){X}}
\put(147,40){\line(1,0){8}}
\put(151,40){\circle*{3}}
\put(147,30){\line(1,0){8}}
\put(151,56){\line(0,-1){2}}
\put(151,46){\line(0,-1){6}}

\put(147,6){\framebox(8,8){Z}}
\put(151,6){\line(0,-1){2}}

\multiput(147,130)(000,-10){13}{\line(-1,0){7}} 

\multiput(132,130)(0,-10){3}{\line(1,0){8}}
\put(132,96){\framebox(8,8){Z}}
\put(132,86){\framebox(8,8){X}}
\put(132,80){\line(1,0){8}}
\put(136,80){\circle*{3}}
\put(132,66){\framebox(8,8){Z}}
\put(136,96){\line(0,-1){2}}
\put(136,86){\line(0,-1){12}}

\put(132,60){\line(1,0){8}}
\put(132,10){\line(1,0){8}}

\put(132,46){\framebox(8,8){Z}}
\put(132,36){\framebox(8,8){X}}
\put(132,30){\line(1,0){8}}
\put(136,30){\circle*{3}}
\put(132,16){\framebox(8,8){Z}}
\put(136,46){\line(0,-1){2}}
\put(136,36){\line(0,-1){12}}

\multiput(122,130)(000,-10){13}{\line(1,0){10}}

\end{picture}

\caption{{\em Left:} The encoding circuit of Fig.\@~\ref{fig:encoding1}
where consecutive blocks of operations have been placed in different
orders and the appropriate commutators introduced. There are $6$ layers of gates in this circuit and in each layer all the gates commute with each other. It is was we call the `pearl-necklace' structure. {\em Right:}
Corresponding decoding circuit obtained by running the modified
encoding circuit backward. In this form it is obvious that the
decoding circuit has a structure allowing a forward
decoding.}\label{fig:decoding1}
\end{figure}
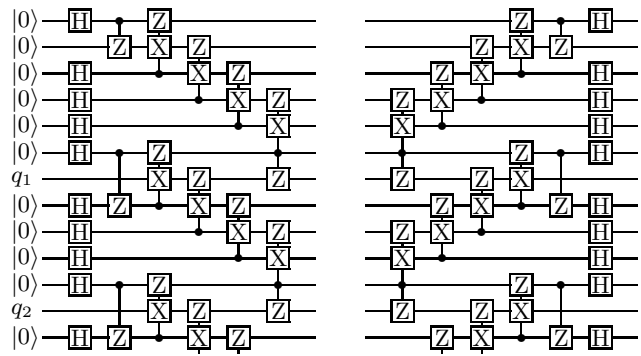

\end{document}